\documentclass[12pt]{article}
\usepackage{graphicx}
\usepackage[utf8]{inputenc}
\usepackage[english
]{babel}
\def\baselinestretch{1.5}
\begin{document}
	\begin{center}
		\bf{Spin kinetic equations in the probability representation of quantum mechanics}\\
	\end{center}
	\bigskip

	\begin{center} {\bf Vladimir N. Chernega$^1$,  Vladimir I. Man'ko$^{1,2}$}
	\end{center}
	
	\medskip
	
	\begin{center}
		$^1$ - {\it Lebedev Physical Institute, Russian Academy of Sciences\\
			Leninskii Prospect 53, Moscow 119991, Russia}\\
		$^2$ - {\it Moscow Institute of Physics and Technology (State University)/
			Institutskii per. 9, Dolgoprudnyi, Moscow Region 141700, Russia }\\	
		Corresponding author e-mail: vchernega\,@\,gmail.com
	\end{center}
	
	\section*{Abstract}
We discuss the possibility to formulate the dynamics of spin states
described by the Schr\"odinger equation for pure states and the von
Neumann equation (as well as the GKSL equation) for mixed states in
the form of quantum kinetic equations for probability distributions.
We review an approach to the spin-state description by means of the
probability distributions of dichotomic random variables.
	\medskip	

\noindent{\bf Keywords:} spin states, kinetic equations,
Gorini--Kossakowski--Sudarshan--Lindblad (GKSL)~equation,
Schr\"odinger equation.
	
\section{Introduction}  
The quantum kinetic equations based on the Schr\"odinger equation
for the wave function~\cite{Sch26} and the von Neumann equation for
the density matrix~\cite{Landau27,vonNeumann27} have been used to
study different physical systems (in particular, the properties of
Fermi liquids have been studied
in~\cite{Silin,Silin1,Silin2,Silin3,Silin4,Silin5,Silin6,Silin7}).
The important role of spin states in the dynamics of such systems as
electron liquids in metals was considered in the presence of
magnetic field in~\cite{Silinmgf}.

The aim of this paper is to review recent approach to the
description of quantum states by means of probability distributions
of standard classical-like random
variables~\cite{Chernega1,Chernega2,Chernega3,Chernega4,Chernega5,Chernega6}.
This approach provides the possibility to construct the map of
density operators and state vectors belonging to a Hilbert space of
quantum-system states onto probability distributions, which means
that all equations, including the Schr\"odinger equation for state
vectors and the von Neumann equation for density operators, can be
mapped onto kinetic equations for the probability distributions.

We present this approach on the example of a system with discrete
variables like the spin-1/2 system and formulate the generic method
for deriving the kinetic equations for classical-like probability
distributions describing any quantum system. Some aspects of such
approach are presented
in~\cite{AvanesovManko,AvanesovManko1,Gusliinen,Kiktenko,MarmoJpAVitaele,ChernegaPhDThesis}.

This paper is organized as follows.

In Sec.~2, the spin-1/2 states are identified with the probability
distribution of dichotomic random variables. In Sec.~3, the quantum
evolution of the spin-1/2 density matrix is presented in the form of
a kinetic equation for dichotomic variables describing the spin
states. In Sec.~4, generic systems with discrete and continuous
variables are considered using the probabilities describing their
states. Conclusions and prospects are given in Sec.~5.

\section{Spin-1/2 Density Matrix}   
The pure state of the spin-1/2 system (e.g., the electron spin in a
metal) is identified with the state vector $|\psi\rangle$, which has
complex components $a$ and $b$ satisfying the normalization
condition $|a|^2+|b|^2=1$. In the case of mixed state of the spin
system, the density 2$\times$2~matrix $\rho$, with complex matrix
elements $\rho_{11}$, $\rho_{12}$, $\rho_{21}=\rho^\ast_{12}$, and
$\rho_{22}$, such that $\rho=\rho^\dag$, $\mbox{Tr}\rho=1$, and
$\rho\geq0$, is identified with the state.

The pure state has the density matrix
$\rho_\psi=|\psi\rangle\langle\psi|$, where the column vector
$|\psi\rangle$ is considered as a rectangular matrix with two rows,
and the vector $\langle\psi|$ is considered as the matrix with two
columns, i.e., $\langle\psi|=(|\psi\rangle)^\dag$. The matrix
$\rho_\psi$ is defined as a 2$\times$2~matrix given by the product
of two rectangular matrices
$\rho_\psi=(|\psi\rangle)(|\psi\rangle)^\dag$.

The evolution equation for the spin system with a Hamiltonian $H$,
which is the 2$\times$2~matrix with matrix elements $H_{11}$,
$H_{12}$, $H_{21}=H_{12}^\ast$, and $H_{22}$, such that $H=H^\dag$,
has the form of the Schr\"odinger equation
\begin{equation} \label{eq.2.1}
i\,\frac{\partial}{\partial t}|\psi(t)\rangle = H(t)|\psi(t)\rangle,
  \qquad \hbar=1.
\end{equation}

The unitary evolution of the state vector, i.e.,
$|\psi(t)\rangle=U(t)|\psi(0)\rangle$, where the unitary
2$\times$2~matrix $U(t)$ has the matrix elements $U_{11}(t)$,
$U_{12}(t)$, $U_{21}(t)$, and $U_{22}(t)$, satisfies the equation
\begin{equation} \label{eq.2.2}
i\,\frac{\partial U(t)}{\partial t} = H(t)U(t), \qquad U(0)=1,
\end{equation}
and one has the unitarity condition $U(t)U^\dag(t)=1$. For the
time-independent Hamiltonian, $U(t)=\exp(-itH)$ and the density
matrix $\rho(t)$ of arbitrary states evolves as
$\rho(t)=U(t)\rho(0)U^\dag(t)$.

We address now the following problem: how to describe the discussed
form of quantum dynamics using standard probability distributions
and their time dependence.

Following~\cite{Chernega1,Vitale,OlgaJETP,DodPLA}, one can check
that, if we denote the matrix elements of the spin density matrix as
$\rho_{11}=p_3$ and $\rho_{12}=p_1-1/2-i(p_2-1/2)$, the real numbers
$0\leq p_1,p_2,p_3\leq1$ have the physical meaning of probabilities
to have spin projections on perpendicular axes $x$, $y$, and $z$
equal to $+1/2$. One can check this fact by calculating the trace
$\mbox{Tr}(\rho\rho_k)$ which, due to Born's rule, is the probability to
have in the state $\rho$ the properties associated with the states
$\rho_k$. If we choose three density matrices $\rho_k$, $k=1,2,3$ as
density matrices
\[
\rho_1 =   \frac{1}{2}\left(\begin{array}{cc}1&1\\1&1\end{array}\right),\qquad
  \rho_2 = \frac{1}{2}\left(\begin{array}{cc}1&-i\\i&1\end{array}\right),\qquad
  \rho_3 = \left(\begin{array}{cc}1&0\\0&0\end{array}\right),
\]
these matrices describe the states with spin projection $+1/2$ onto
the directions $x$, $y$, and $z$, since they correspond to
eigenvectors $|\psi_1\rangle$, $|\psi_2\rangle$, and
$|\psi_3\rangle$ of the Pauli matrices $\sigma_1$, $\sigma_2$, and
$\sigma_3$; the obtained numbers $p_1$, $p_2$, and $p_3$ have
exactly the meaning of probabilities. These three numbers can be
used to get all the matrix elements of the density matrix $\rho$
expressed as linear combinations of the probabilities.

Thus, we obtain an invertible map $\rho\leftrightarrow
p_1,~p_2,~p_3$ of the density matrix onto three probability
distributions $(p_1,1-p_1)$, $(p_2,1-p_2)$, and $(p_3,1-p_3)$ of
dichotomic random variables. The probability distributions can be
interpreted as probabilities describing statistics of three nonideal
classical coins.

For classical coins that are independent (there is no correlations),
the numbers $p_1$, $p_2$, and $p_3$ satisfy only one condition
$~0\leq p_1,p_2,p_3\leq 1$.

For spin states, the nonnegativity of the density matrix provides
the constraint $(p_1-1/2)^2+(p_2-1/2)^2+(p_3-1/2)^2\leq1/4$ that
corresponds to the presence of quantum correlations of the
spin-projections onto three different directions.

As it was always considered before, the quantum nature of the system
behavior (like the spin-1/2 system) needed the formalism of Hilbert
space vectors and operators acting in this space to develop the
theory of physical phenomena. But it turned out that, as in
classical theory, it is sufficient to use the formalism of standard
probability theory to describe the states of the systems, e.g., the
spin-1/2 system. The only ingredient introduced by the quantum
mechanics formalism is the fact that in nature there exist quantum
correlations, which impose extra constraints onto the probability
distributions expressed in terms of inequalities, which do not exist
in corresponding classical systems like three coins.

The three probability distributions can be considered as conditional
probability distributions $P(m|j)$, $m=\pm1/2$, $j=1,2,3$; here,
$P(1/2|1)=p_1$, $P(1/2|2)=p_2$, and $P(1/2|3)=p_3$. The joint
probability distribution of two random variables ${\cal P}(m,j)$
provides three conditional probability distributions due to Bayes'
formula ${\mathcal P}(m,j)=P(m|j)\left(\sum_{m=-1/2}^{1/2}{\mathcal
P}(m,j)\right)$. Here, $\sum_{m=-1/2}^{1/2}{\mathcal P}(m,j)$ is the
marginal probability distribution to have a random variable $j$,
which plays the role of the coin number in the classical case ---
the number $P(m|j)$ is the probability of the spin projection $m$ on
the direction of the axes $x$, $y$, and $z$ in the experiments where
the spin projections $+1/2$ onto the three directions are measured.

A simple example of such marginal probability distribution is
$\sum_{m=-1/2}^{1/2}{\mathcal P}(m,j)=\left(1/3,\,1/3,\,1/3\right)$.
In this case, the joint probability distribution ${\mathcal P}(m,j)$
contains six probabilities expressed by the probability 6-vector
$\vec{\mathcal
P}=\frac{1}{3}\left(p_1,\,1-p_1,\,p_2,\,1-p_2,\,p_3,\,1-p_3\right)$.

\section{Evolution of the Spin Quantum State as the Time Dependence of
the Probability Distribution}   
The unitary evolution equation
for the density matrix of the spin-1/2 system, associated with the
Hamiltonian $H$, which is a Hermitian 2$\times$2~matrix with matrix
elements $H_{j k}$, $j,k=1,2$, reads
\[
i\,\frac{\partial\rho}{\partial t} = H\rho-\rho H;
\]
it is the kinetic equation for the probabilities $p_1(t)$, $p_2(t)$,
and $p_3(t)$ of the form
\begin{equation} \label{eq.K1}
i\left(\begin{array}{cc}
  \dot p_3&\dot p_1-i\dot p_2\\
  \dot p_1+i\dot p_2&-\dot p_3
\end{array}\right)=\left[\left(
  \begin{array}{cc}H_{11}&H_{12}\\H_{21}&H_{22}\end{array}\right),
\left(\begin{array}{cc}p_3&p_1-1/2-i(p_2-1/2)\\
  p_1-1/2+i(p_2-1/2)&1-p_3\end{array}\right)\right].
\end{equation}
This von Neumann equation for the density matrix is equivalent to a
linear differential equation for the probability 6-vector
$\vec{{\cal P}}(t)$ of the form $\frac{d\vec{{\cal P}}}{d
t}=\hat H\vec{\mathcal P}+\vec\Gamma$. The matrix elements of the
6$\times$6~matrix $\tilde{H}$ and numerical 6-vector $\vec\Gamma$ in
this kinetic equation can be easily obtained from Eq.~(\ref{eq.K1});
see~\cite{AvanesovManko}.

The solution to the kinetic equation~(\ref{eq.K1}) provides the
linear transform of the vector $\vec{\mathcal{P}}$ given by the
relation
\begin{eqnarray}    \label{eq.K2}
\left(\begin{array}{cc}
  p_3(t)& p_1(t)-1/2-i(p_2(t)-1/2)\\
  p_1(t)-1/2+i(p_2(t)-1/2)&1-p_3(t)
  \end{array}\right) =
\left(\begin{array}{cc}U_{11}(t)&U_{12}(t)\\U_{21}(t)&U_{22}(t)\end{array}\right)\nonumber\\
\times
\left(\begin{array}{cc}p_3(0)&p_1(0)-1/2-i(p_2(0)-1/2)\\p_1(0)-1/2+i(p_2(0)-1/2)&1
-p_3(0)\end{array}\right)\left(\begin{array}{cc}U_{11}^\ast(t)&H_{21}^\ast(t)\\
H_{12}^\ast(t)&H_{22}^\ast(t)\end{array}\right),
\label{eq.K2}
\end{eqnarray}
with the unitary matrix $U(t)=\exp(-i t H)$. For an arbitrary
unitary matrix $U(t)$, the time dependence of probabilities
$p_1(t)$, $p_2(t)$, and $p_3(t)$ provides the trajectory of the
probability distribution on the simplex, which respects the
constraints corresponding to quantum correlations of spin
projections on the perpendicular directions given by axes $x$, $y$,
and $z$.

The 2$\times$2~matrix $\rho$ can be mapped onto a column 4-vector
$\vec\rho$ with four components $\rho_1=\rho_{11}$,
$\rho_2=\rho_{12}$, $\rho_3=\rho_{21}$, and $\rho_4=\rho_{22}$. In
view of the map described, the unitary evolution of the matrix
$\rho(t)\rightarrow u(t)\rho(0)u^\dag(t)$ can be presented as the
evolution of the vector $\vec\rho$; namely, the 4-vector
$\vec\rho(t)=u(t)\otimes u^\ast(t)\vec\rho(0)$. The
4$\times$4~matrix $u(t)\otimes u^\ast(t)$ is a unitary matrix; it
transforms the initial probabilities providing the map of the
4-vector $\vec{\mathcal{P}}(0)$ with components $p_3(0)$,
$p_1(0)-(1/2)-i[p_2(0)-(1/2)]$, and $p_1(0)-(1/2)+i[p_2(0)-(1/2)]$,
$1\to p_3(0)$ onto a vector $\vec{\mathcal{P}}(t)$, namely,
$\vec{\mathcal{P}}(t)=u(t)\otimes u^\ast(t)\vec{\mathcal{P}}(0)$.
Here, vectors $\vec{\mathcal{P}}(t)$ and $\vec{\mathcal{P}}(0)$ are
columns with vector components expressed in terms of probabilities
describing the qubit-state density matrix.

We also can get the evolution of such vectors, which are associated
with arbitrary quantum channels. In this case, the evolution of the
vector with probability components is given by an analogous
expression, where the unitary matrix $u(t)\otimes u^\ast(t)$ is
replaced by the 4$\times$4~matrix $V(t)$ of the form associated with
the Sudarshan--Mathews--Rau--Kraus
transform~\cite{SudarshanMathewsRaoPR,Krausbook}
$V(t)=\sum_k\big(S_k(t)\otimes S_k^\ast(t)\big)$. Here, $S_k$ are
arbitrary 2$\times$2~matrices satisfying the condition $\sum_k
S_k^\dag S_k=1_2$.

For example, if matrices $S_k(t)$ are such that
$S_k(t)=\sqrt\lambda_k u_k(t)$, where $u_k(t)$ are unitary
2$\times$2~matrices and numbers $\lambda_k$ such that
$0\leq\lambda_k\leq1$ satisfy the normalization condition
$\sum_k\lambda_l=1$, the transformation of probability vectors
$\vec{\mathcal{P}}(0)\rightarrow\vec{\mathcal{P}}(t)$ gives the
pseudostochastic map~\cite{Gusliinen} of the 6-vector with the
components
$\frac{1}{3}\,\big(p_1(0),\,1-p_1(0),\,p_2(0),\,1-p_2(0),\,p_3(0),\,1-p_3(0)\big)$
corresponding to the initial density matrix of the qubit state
$\rho(0)$.

The probability 6-vector $\vec{\mathcal{P}}(t)$ with components
$\frac{1}{3}\,\big(p_1(t),\,1-p_1(t),\,p_2(t),\,1-p_2(t),\,p_3(t),\,1-p_3(t)\big)$
satisfies the linear kinetic equation determined by the GKSL
equation. Thus, we can obtain an arbitrary quantum evolution
equation like the von Neumann equation or the
Gorini--Kossakowski--Sudarshan--Lindblad
(GKSL)~equation~\cite{Gorini,Lindblad} in the form of a kinetic
equation for the probability distribution determining quantum
states. This approach can be extended to arbitrary spin states as
well as to the description of systems with continuous variables like
a quantum parametric oscillator.

In the probability representation, the oscillator states are
described by quantum symplectic
tomograms~\cite{ManciniTombesi,Man1JRLR}. Also these states can be
described by optical tomograms, which are the probability
distributions satisfying the evolution equation in the form of the
kinetic equation~\cite{KorennoyJRLR,AmosovKorennoyMankoPhysRev}.

\section{General Case of Qudits}   
The density $N$$\times$$N$~matrix of the qudit state ($N$-level
atom, spin-$j$ state with $2j+1=N$) has the matrix elements $\rho_{j
k}$, $j,k=1,2,\ldots N$, for which $\rho_{j k}^\ast=\rho_{k j}$,
$\sum_k\rho_{k k}=1$, and $\rho\geq0$, i.e., eigenvectors of the
matrix are nonnegative. It was
found~\cite{Chernega1,Chernega2,Chernega3,MarmoJpAVitaele,Vitale}
that the matrix elements of this matrix are expressed in terms of
probabilities $0\leq p^{(j k)}_{1,2,3}\leq1$ of dichotomic random
variables such that
\begin{eqnarray}
&&\rho_{jk} = p_1^{(jk)}-(1/2)-i(p_2^{(j k)}-1/2),\qquad j<k, \nonumber\\
&&\rho_{jj}=p_3^{jj},\quad j=1,2,\ldots,N-1, \label{eq.K11}\\
&&\rho_{N N}=1-\sum_{j=1}^{N-1}p_3^{j j}.  \nonumber
\end{eqnarray}
There exist other probability distributions for the density
matrix~\cite{MarmoJpAVitaele,Vitale}, which depend linearly on the
probabilities $p_{1,2,3}^{(jk)}$. All properties of the
probabilities determining qubit states are valid also for qudit
states and for any states with $N=1,2,\ldots,\infty$; for example,
for the density matrix $\rho_{jk}$ of harmonic oscillator written in
the Fock basis. In this case, we have an infinite number of
probability distributions $0\leq p_{1,2,3}^{(jk)}\leq1$ of
dichotomic random variables determining the oscillator quantum
state. The kinetic equation for the probability distribution has the
form of a von Neumann equation for the oscillator density matrix in
the position representation; it reads
\begin{equation} \label{eq.K12}
i\frac{\partial\rho(x,x',t}{\partial t} = -\frac{1}{2}
   \left(\frac{\partial^2}{\partial x^2}
  -\frac{\partial^2}{\partial {x'}^2}\right)\rho(x,x',t)
  +\frac{x^2-{x'}^2}{2}\rho(x,x',t), \qquad \hbar = m = \omega = 1.
\end{equation}
The density matrix in the Fock basis $\rho_{jk}=\langle j|\hat
\rho| k\rangle$ is connected with the matrix in the position representation
$\langle x|\hat\rho|x'\rangle$ as follows:
\[
\rho(x,x') = \sum_{k,j=0}^\infty\langle x|k\rangle\langle k|
  \hat\rho|j\rangle\langle j|x'\rangle,\qquad
  k,j=0,1,2,\ldots,\infty;
\]
this means that the density matrix $\rho(x,x',t)$ satisfying
Eq.~(\ref{eq.K12}) is expressed in terms of probabilities
$p_{1,2,3}^{(jk)}$ as
\begin{equation} \label{eq.K14}
\rho(x,x',t) =
\sum_{k,j=0}^\infty\frac{e^{\left(x^2+{x'}^2\right)/2}
  }{\sqrt\pi}\frac{H_k(x)H_j(x')}{\sqrt{k!j!2^{k+j}}}\rho_{kj}(t),
\end{equation}
where the matrix elements
\begin{equation} \label{eq.K15}
\rho_{kj}(t) = p_1^{(kj)}(t)-1/2-i\left(p_2^{(k,j)}(t)-1/2\right),\qquad
 k<j,\qquad \rho_{kk}(t)=p_3^{(kk)}(t)
\end{equation}
are expressed in terms of probabilities $p_{1,2,3}^{(jk)}$
satisfying the kinetic equation corresponding to the used quantum
evolution equation either for the unitary or nonunitary evolution.

\section{Conclusions}    
The authors of this paper have the privilege to discuss with Prof.
V.~P.~Silin the problem of pro\-bability representation of quantum
evolution equations, especially in connection with the PhD Thesis of
V.~N.~Chernega~\cite{ChernegaPhDThesis}.  Professor V.~P.~Silin
actively participated in the discussion during V.~N.~Chernega's
defence of his PhD Thesis at the Scientific Council of the Lebedev
Physical Institute where Prof.~V.~P.~Silin was a permanent member.

Professor V. P. Silin pointed out that the Wigner representation of
the density matrix and the kinetic equation for quasidistributions
are used in plasma physics as well as in considering the
quantum-liquid
properties~\cite{Silin1,Silin2,Silin3,Silin4,Silin5,Silin6} since it
was useful namely due to its applications to the important physical
effects. In this connection, Prof. V. P. Silin's advice was to apply
the approach based on the probability representation of the
quantum-state density matrix for both discrete variables like spin
projections and for continuous variables like the particle's
position and momenta to clarify the advantages of the new approach.
New aspects of this approach that we hope to use are the
information-entropic characteristics of the density-matrix elements.
These characteristics can be expressed in terms of inequalities for
the Shannon entropy~\cite{Shannon} like the subadditivity condition
existing for different probability distributions. These aspects of
solutions of the kinetic equation determining quantum states will be
considered in future publications. We are grateful to Prof.
V.~P.~Silin for fruitful discussion and advices, and we dedicate
this paper to the memory of Prof. V.~P.~Silin, a great scientist.

\end{document}